\documentclass[12pt]{article}
\usepackage{graphicx}

\usepackage[bookmarksnumbered,plainpages]{hyperref}

\usepackage[format=hang,labelfont=bf,font={small,sf},margin=10pt]{caption}

\begin{document}
\title{Path probability distribution of stochastic motion of non dissipative systems: a classical analog of Feynman factor of path integral}
\author{T.L. Lin$^\dag$$^\flat$$^\S$, R. Wang$^\ddag$, W.P. Bi$^\flat$, \\ A. El Kaabouchi$^\dag$, C. Pujos$^\dag$, F. Calvayrac$^\flat$ \\ Q.A. Wang$^\dag$$^\flat$\thanks{Corresponding author: awang@ismans.fr}\\
$^\dag$LUNAM Universit\'{e}, ISMANS, Laboratoire de Physique Statistique et \\ 
Syst\`emes Complexes, 44, Avenue F.A. Bartholdi, Le Mans, France\\
$^\flat$Facult\'e des Sciences et Techniques, Universit\'{e} du Maine, \\ 
Ave. O. Messiaen, 72035 Le Mans, France\\
$^\S$Department of Physics, Xiamen University, Xiamen, 361005, China\\
$^\ddag$Center for Complex Networks and Systems Research,\\ 
School of Informatics and Computing, Indiana University, USA}

\date{}

\maketitle

\begin{abstract}
We investigate, by numerical simulation, the path probability of non dissipative mechanical systems undergoing stochastic motion. The aim is to search for the relationship between this probability and the usual mechanical action.
The model of simulation is a one-dimensional particle subject to conservative force and Gaussian random displacement. The probability that a sample path between two fixed points is taken is computed from the number of particles moving along this path, an output of the simulation, devided by the total number of particles arriving at the final point. It is found that the path probability decays exponentially with increasing action of the sample paths. The decay rate increases with decreasing randomness. This result supports the existence of a classical analog of the Feynman factor in the path integral formulation of quantum mechanics for Hamiltonian systems.
\end{abstract}


\vfill

\section{Introduction}

The path (trajectory) of stochastic dynamics in mechanics has much richer physics content than that of the regular or deterministic motion. A path of regular motion always has probability one once it is determined by the equation of motion and the boundary conditions, while a random motion may have many possible paths under the same conditions, as can be easily verified with any stochastic process\cite{Mazo}. For a given process between two given states (or configuration points with given durations), each of those potential paths has some chance (probability) to be followed. The path probability is a very important quantity for understanding and characterizing random dynamics because it contains all the information about the physics: the characteristics of the stochasticity, the degree of randomness, the dynamical uncertainty, the equations of motion and so forth. Consideration of paths has long been regarded as a powerful approach to non equilibrium thermodynamics\cite{Onsager,Freidlin,Touchette,Maier,Aurell,Roma,Cohen,JWang,Poulomi,Stock,Fujisaki,Evans,Abaimov}. A key question in this approach is what are the random variables which determine the probability. The Onsager-Machlup type action\cite{Onsager,Maier,Aurell,Roma,Cohen,JWang,Poulomi,Stock,Fujisaki} is one of the answers for Gaussian irreversible process close to equilibrium where the path probability is an exponentially decreasing function of the action calculated along thermodynamic paths in general\cite{Onsager}. This action has been extended to Cartesian space in \cite{Fujisaki}. The large deviation theory\cite{Freidlin,Touchette} suggests a rate function to characterize an exponential path probability. There are other suggestions by the consideration of the energy along the paths\cite{Evans,Abaimov}. For a Markovian process with Gaussian noises, the Wiener path measure\cite{Mazo} provides a good description of the path likelihood with the product of Gaussian distributions of the random variables.

The question we want to answer in this work is the following: suppose a mechanical random motion is trackable, i.e., the mechanical quantities of the motion under consideration such as position, velocity, mechanical energy and so on can be calculated with certain precision along the paths, is it possible to use the time cumulation, along the paths, of Hamiltonian or Lagrangian (action) to characterize the path probability of that motion? Possible answers have been given in \cite{Evans,Abaimov}. The author of \cite{Evans} suggests that the path probability decreases exponentially with increasing average energy along the paths\cite{Evans}. This theory risks a conflict with the regular mechanical motion in the limit of vanishing randomness because the surviving path would be the path of least average energy, while it is actually the path of least action. The proposition of \cite{Abaimov} is a path probability decreasing exponentially with the sum of the successive energy differences, which risks the similar conflicts with regular mechanics mentioned above.

Another proposition, free from the above mentioned conflicts, is to relate the path probability to action, the key quantity for determining paths of Hamiltonian systems in classical mechanics. According to a theoretical work in \cite{Wang2,Wang3,Wang4}, for the special case of Hamiltonian system conserving statistically its energy, the path probability can be distributed in exponential function $e^{-\gamma A}$ of the action $A=\int (K-V)dt$ where $K$ is the kinetic energy, $V$ the potential energy, $\gamma$ a characteristic parameter of the random dynamics and the time integral is carried out along the considered path. This distribution function is analogous to the Feynman factor $e^{\frac{i}{\hbar}A}$ of quantum mechanics\cite{Feynman}. The Feynman factor is not a probability, but in the presence of the quantum randomness, the action $A$ indeed characterizes the way the system evolves along the configuration paths from one quantum state to another\cite{Pavon}. In both (classical\cite{Wang2,Wang3,Wang4} and quantum\cite{Feynman}) versions, the system statistically remains Hamiltonian in spite of the classical or quantum randomness. The classical path is recovered when the randomness is vanishing with infinite $\gamma$ or zero Planck constant $\hbar$. 

The aim of this work is to check this prediction for classical mechanics by means of numerical simulation of the random motion of Hamiltonian systems. There are several reasons for limiting this work to Hamiltonian systems. Firstly, action is only well defined for Hamiltonian (often energy conservative) mechanical systems\cite{Lanczos,Sieniutycz}. Secondly, from the previous results \cite{Onsager}-\cite{Abaimov} for diffusive and random motion (usually non conservative systems), the paths do not simply depend on the usual action, in general. Thirdly, the random motion without dissipation is certainly an ideal model, but it is also a good approximation to many real random motions which are weakly damped with negligible energy dissipation compared to the variation of potential energy. These are the cases where the conservative force is much larger than the friction ones. In other words, the system is (statistically) governed by the conservative forces. These motions are frequently observed in Nature. We can imagine, e.g., a falling motion of a particle which is sufficiently heavy to fall in a medium with acceleration approximately determined by the conservative force during a limited time period, but not too heavy in order to undergo observable randomness due to the collision from the molecules around it or to other sources of randomness. Other examples include the frequently used ideal models of thermodynamic processes, such as the free expansion of isolated ideal gas and the heat conduction within perfectly isolated systems which, in spite of the thermal fluctuation, statistically conserve energy during the motion. Hence the result of this work is expected not only to answer a fundamental question concerning path probability and action, but also to be useful as a mathematical tool for investigating some real dynamics.

This ideal model can be depicted as the following Langevin equation 
\begin{equation} \label{0}
m\frac{d^2x}{dt^2}=-\frac{dV(x)}{dx}-m\zeta\frac{dx}{dt}+R
\end{equation}
with the zero friction limit (friction coefficient $\zeta\rightarrow 0$), where $x$ is the one dimensional position, $t$ the time, $V(x)$ the potential energy and $R$ the Gaussian distributed random force. For this motion, a stochastic Hamiltonian/Lagrangian mechanics has been formulated in \cite{Wang2,Wang3,Wang4} where a path entropy is introduced to measure the dynamical randomness or uncertainty in the path probability distribution. When this path entropy takes the form of the Shannon formula, the maximum entropy calculus stemming from an stochastic version of least action principle leads to an exponentially decreasing path probability with increasing action. The numerical simulation of this work to verify this theoretical prediction can be summarized as follows. We track the motion of a large number of particles subject to a conservative force and a Gaussian distributed random displacement. The number of particles from one given position to another through some sample paths is counted. When the total number of particles are sufficiently large, the probability (or its density) of a given path is calculated by dividing the number of particles counted along this path by the total number of particles arriving at the end point through all the sample paths. The correlation of this probability distribution with two mechanical quantities, the action and the time integral of Hamiltonian calculated along the sample paths, is analyzed. In what follows, we first give a detailed description of the simulation, followed by the analysis of the results and the conclusion.

\section{Technical details of numerical computation}

The numerical model of the random motion can be outlined as follows. The particles are subject to a conservative force and a Gaussian noise (random displacements $\chi$, see Eq.(\ref{1}) below) and move along the axis $x$ from an initial point $a$ (position $x_0$) to a final point $b$ ($x_n$) over a given period of time $n\delta t$ where $n$ is the total number of discrete steps and $\delta t=t_i-t_{i-1}$ the time increment of a step which is the same for every step. Many different paths are possible, each one being a sequence of random positions $\{x_{0},x_{1},x_{2}\cdots x_{n-1},x_{n}\}$, where $x_i$ is the position at time $t_i$ ($i=0,1,2 ...n$) and generated from a discrete time solution of Eq.(\ref{0}) :
\begin{equation} \label{2}
x_{i}=x_{i-1}+\chi_i+f(t_i)-f(t_{i-1}),
\end{equation}
which is a superposition of a Gaussian random displacement $\chi_i$ and a regular motion $y_i=f(t_i)$, the solution of the Newtonian equation $m\frac{d^2x}{dt^2}=-\frac{dV(x)}{dx}$ corresponding to the least action path. Naturally, Eq.(\ref{2}) is a solution of Eq.(\ref{0}) under the condition that the superposition principle is valid for this motion. This principle should work whenever Eq.(\ref{0}) is linear, for instance, without force, or with constant and harmonic forces. The reader will find that we have also used two others forces which make Eq.(\ref{0}) nonlinear and may invalidate the superposition property of Eq.(\ref{2}). Nevertheless linear equation is sufficient but not necessary for superposition. From the fact that the results from these two potentials are similar to those from linear equation, the superposition seems to work, at least approximately. We think that this positive result with nonlinear forces may be attributed to two favorable elements : 1) most of the random displacements per step are small compared to the regular displacement; 2) the random nature of these small displacements may statistically cancel the nonlinear deviation from superposition property.

For each simulation, we select about 100 sample paths randomly created around the least action path $y=f(t)$. The magnitude of the Gaussian random displacements is controlled to ensure that all the sample paths are sufficiently smooth but sufficiently different from each other to give distinct values of action and energy integral. 

A width $\delta$ is given to each sample path which becomes a smooth tube with the axial line composed of a sequence of positions $\{z_{0},z_{1},z_{2}\cdots z_{n-1},z_{n}\}$. The reason for this is the following. In principle, the probability of a single path (a geometrical line $\{x_{0},x_{1},x_{2}\cdots x_{n-1},x_{n}\}$ with zero thickness) is vanishingly small. Only its probability density is meaningful, as discussed later in the conclusion. We should specify here that, in practice, with the limited number of particles in a simulation and the precision of the position, there are hardly more than one particle moving along a same geometrical line $\{x_{0},x_{1},x_{2}\cdots x_{n-1},x_{n}\}$. Typically we have one particle along one geometrical line saved in the output of the simulation. 

To calculate the probability density, a sample path must be defined as a tube of finite thickness $\delta$ (a band in the $x-t$ representation). As all the particles or their trajectories from $a$ to $b$ are saved in the simulation, the number $N_k$ of particles (or geometrical lines) going through a given sample path or tube $k$, i.e., all the sequences of positions $\{x_{0},x_{1},x_{2}\cdots x_{n-1},x_{n}\}$ satisfying $\{z_i\}_k-\delta/2\leq x_i\leq \{z_i\}_k+\delta/2$ for all $i=1,2 ...n$, can be counted, $\{z_{0},z_{1},z_{2}\cdots z_{n-1},z_{n}\}_k$ being the axial line of that tube. The probability that the path $k$ is taken is given by $P_k=N_k/N$ where $N$ is the total number of particles (trajectories) moving from $a$ to $b$ through all the considered sample paths. $P_k$ for each sample path fluctuates a lot from one simulation to another when the total number of particles simulated is small, and tends to a stable value when we gradually increase the particle number. The largest number of particles we used is $10^9$ at which the value of $P_k$ for a given sample path does not change significantly even the number of particles is increased further.

The probability density $\rho_k$ of the sample path $k$ is defined by $\rho_k=\frac{p_k}{\delta^n}$ for a path of $n$ steps. In this paper, the value of $P_k$ is used everywhere for the sake of simplicity.

Some words about the thickness $\delta$ of the sample paths. $\delta$ must be sufficiently large in order to include a considerable number of trajectories in each tube for the calculation of reliable path probability, but sufficiently small in order that the positions $z_i$ and the instantaneous velocities $v_i$ determined along an axial line be representative of all the trajectories in a tube. If $\delta$ is too small, there will be few particles going through each tube, making the calculated probability too uncertain. If it is too large, $z_i$ and $v_i$, as well as the energy and action of the axial line will not be enough representative of all the trajectories in the tube. The $\delta$ used in this work is chosen to be 1/2 of the standard deviation $\sigma$ of the Gaussian distribution of random displacements. The left panels of figures 1-5 illustrate the axial lines of the sample paths. 

For each sample path, the instantaneous velocity at the step $i$ is calculated by $v_i=\frac{z_i-z_{i-1}}{t_i-t_{i-1}}$ along the axial line. This velocity can be approximately considered as the average velocity of all the trajectories passing through the tube. The kinetic energy is given by $K_i=\frac{1}{2}mv_i^2$, the action by $A_L=\sum_{i=1}^{10} [\frac{1}{2}mv_i^2-V(x_i)]\cdot \delta t$, called Langrangian action from now on in order to compare with the time integral of Hamiltonian $A_H=\sum_{i=1}^{10} [\frac{1}{2}mv_i^2+V(x_i)]\cdot \delta t$ referred to as Hamiltonian action. The magnitude of the random displacements and the conservative forces are chosen such that the kinetic and potential energy are of the same order of magnitude. This allows to clearly distinguish the two actions along a same path. The path probability will be plotted versus $A_L$ and $A_H$ as shown in the Figures 1-5. 

In order to simulate a Gaussian process close to a realistic situation, we chose a spherical particle of 1-$\mu$m-diameter and of mass $m=1.39\times10^{-15}$ kg. Its random displacement at the step $i$ is produced with the Gaussian distribution
\begin{equation} \label{1}
p(\chi_i,t_i-t_{i-1})=\frac{1}{\sqrt{2\pi}\sigma} e^{-\frac{\chi_i^2}{2\sigma^2}},
\end{equation}
where $\chi_i$ is the Gaussian displacement at the step $i$, $\sigma=\sqrt{2D(t_i-t_{i-1})}=\sqrt{2D\delta t}$ the standard deviation, $D=\frac{k_BT}{6\pi r\eta}$ the diffusion constant, $k_B$ the Boltzmann constant, $T$ the absolute temperature, and $r=0.5$ $\mu$m the radius of the particle. For the viscosity $\eta$, we choose the value $8.5\times10^{-4}$ $Pas$ of water at room temperature\footnote{Note that this viscosity is chosen to create a realistic noise felt by the particle as if it was in water. But this viscosity and the concomitant friction do not enter into the equation of motion Eq.(\ref{2}).}. In this case, $D\approx4.3\times10^{-13}$ $m^2/s$ and $\sigma\approx 3\times 10^{-9}$ $m$ with $\delta t=10^{-5}$ $s$. The relaxation time is close to $10^{-7}$ $s$. With this reference, the simulations were made with different time increments $\delta t$ ranging from $10^{-7}$ to $10^{-3}$ $s$. Due to the limited computation time, we have chosen $n=10$. 

We would like to emphasize that the simulation result should be independent of the choice of the particle size, mass, and water viscosity etc. For instance, if a larger body is chosen, the magnitude ($\sigma$) of the random displacements and the time duration of each step, will be proportionally increased in order that the paths between two given points are sufficiently different from each other. 

In what follows, we will describe the results of the numerical experiments performed with 5 potential energies: free particles with $V(x)=0$, constant force with $V(x)=mgx$, harmonic force with $V(x)=\frac{1}{2}kx^2$ and two other higher order potentials $V(x)=\frac{1}{3}Cx^3$ and $V(x)=\frac{1}{4}Cx^4$ $(C>0)$ to check the generality of the results.

\section{Path probability by numerical simulation}

As mentioned above, in each numerical experiment, we launch up to $10^9$ particles from the initial point $a$. Several thousands $N$ arrive, passing by all the sample paths, at the destination $b$ in the interval $z_{b}-\delta/2\leq x_b\leq z_{b}+\delta/2$. The output of the simulation is $N$ and $N_k$ for every sample path whose actions have been already calculated. Once the path probability is determined by $p_k=N_k/N$, its correlation with the Lagrangian or Hamiltonian action of the sample paths can be found by drawing the probability values against the two actions. With $10^9$ particles launched at $a$, the calculated probability values are quite reliable, in the sense that more particles and longer computation time do not produce remarkable improvement of the probability distribution of action. The results presented below for each potential were obtained with $\delta t=10^{-5}$ $s$.

\subsection{Free particles}

Free particles have zero potential energy and constant $f(t)$. So these is no difference between the Lagrangian and Hamiltonian actions. The right panel of Figure 1 shows a path probability which can be well described by the following exponential function
\begin{equation}    \label{5}
P(A_k)=\frac{1}{Z} e^{-\gamma A_k},
\end{equation}
where $A_k$ is either the Lagrangian or Hamiltonian action of the path $k$, and $\gamma\approx 6.7\times10^{26}$ $J^{-1}s^{-1}$. The normalization function $Z$ can be analytically determined by the path integral technique\cite{Feynman}
\begin{equation}
\int_{-\infty}^{\infty}P(A_k)\prod_{i=1}^{n-1}(\frac{dx_{i}}{\delta})=1
\end{equation}
with fixed $x_a$ and $x_b$, or numerically by the value of $\ln p(A=0)$ which can be found with the distribution curves in the figures.

\begin{figure} \label{fig:1}
\begin{minipage}[t]{0.45\linewidth}
\centering
\includegraphics[width=2.5 in]{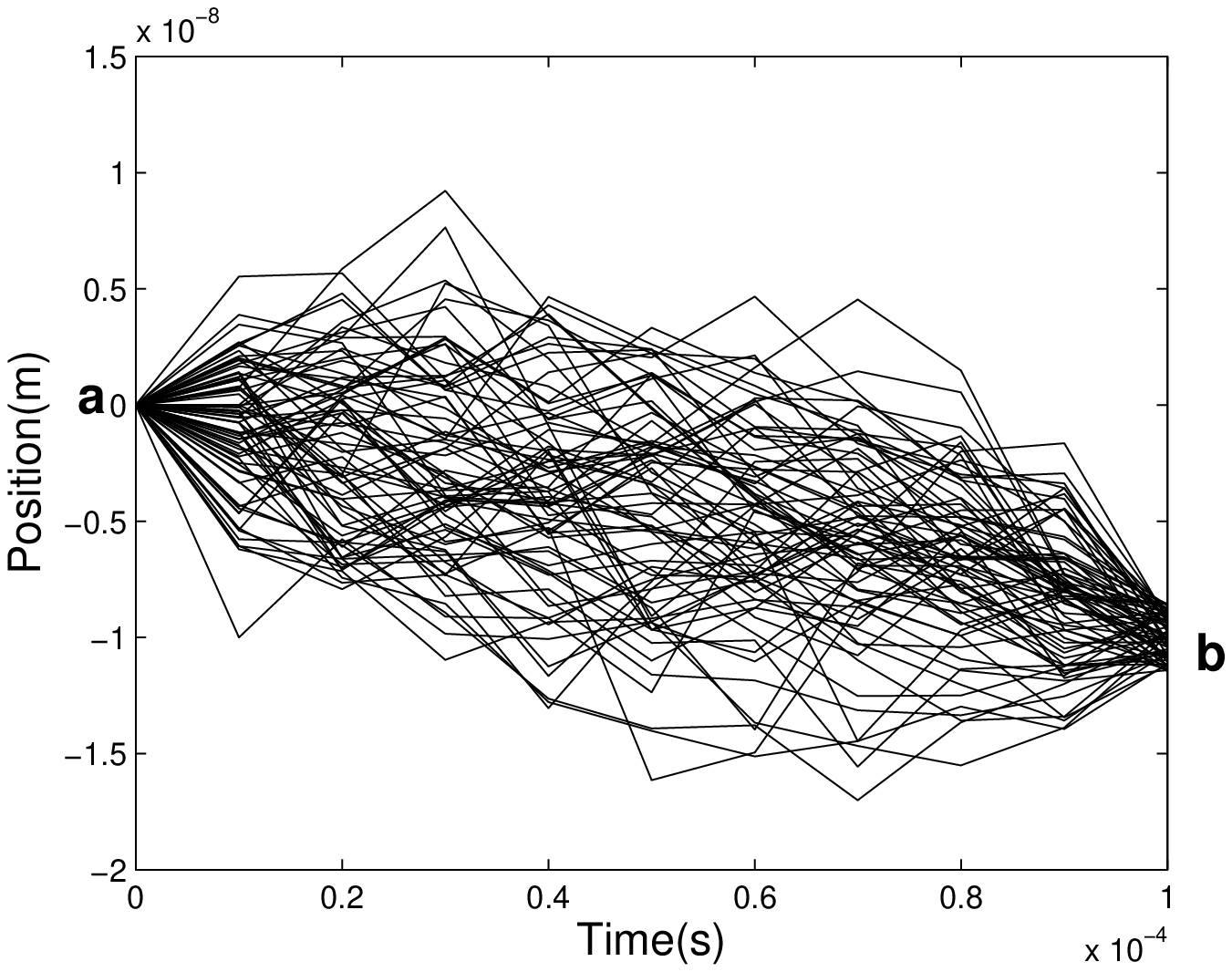}
\end{minipage}
\begin{minipage}[t]{0.45\linewidth}
\centering
\includegraphics[width=2.5 in]{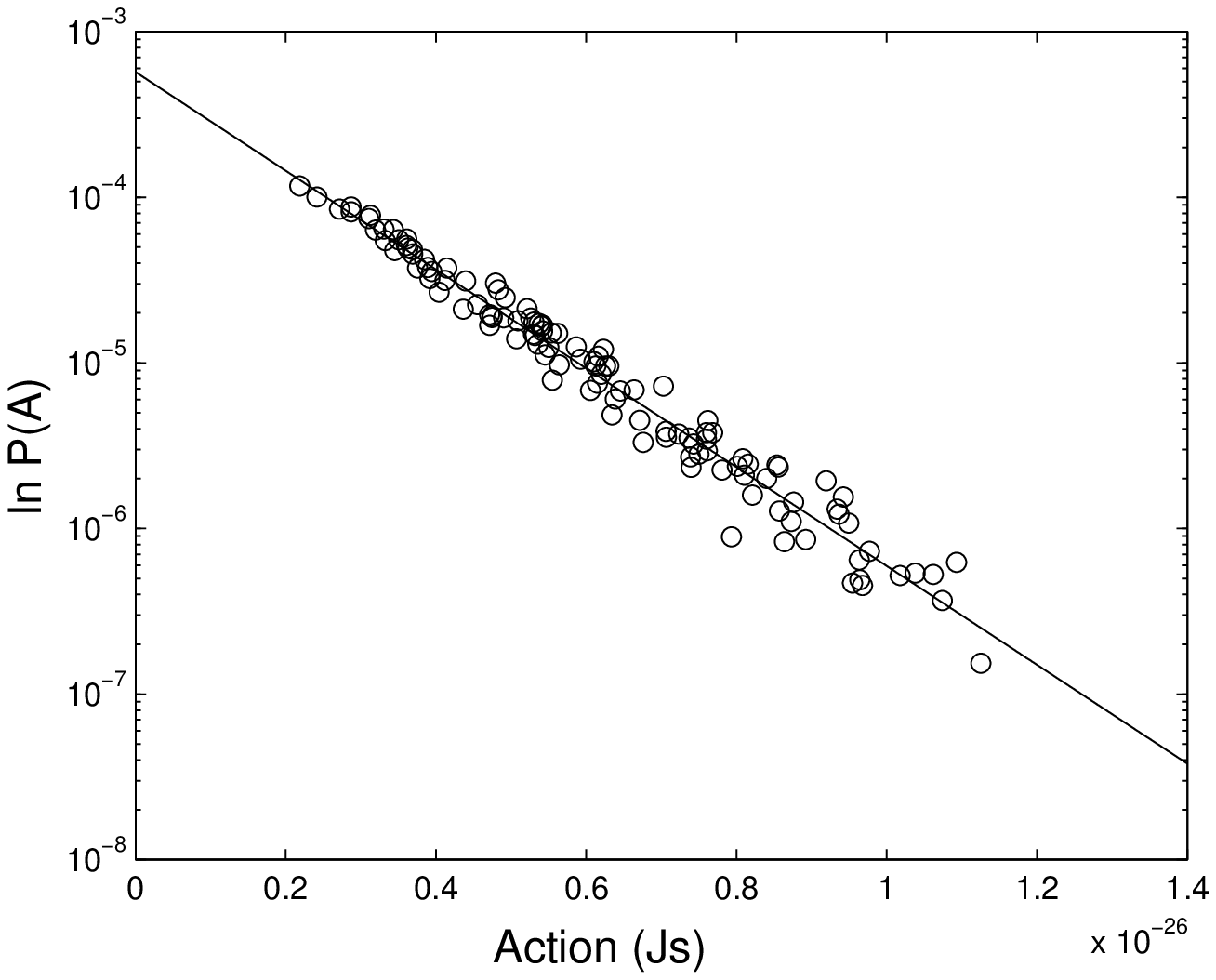}
\end{minipage}
\caption{Result for free particles. The left panel shows the axial lines of the sample paths between the given points $a$ and $b$. The right panel shows the path probability distribution against the Lagrangian and Hamiltonian actions which are equal here as $V(x)=0$. The straight line is a best fit of the points with a slope of about $~-6.7\times10^{26}$ $J^{-1}s^{-1}$.}
\end{figure}

\subsection{Particles under constant force}

To distinguish the dependences of the path probability on Lagrangian and Hamiltonian actions, it is necessary to see random motion under conservative forces. The first force we studied is the constant force with the potential $V(x)=mgx$. The regular motion is described by  $f(t)=-\frac{1}{2}gt^2$, where the parameter $g=10 m/s^2$. The results are shown in the right panel of Figure 2. Eq.(\ref{5}) still holds with $\gamma\approx 6.4\times10^{26}$ $J^{-1}s^{-1}$. There is no correlation between path probability and Hamiltonian action.

\begin{figure} \label{fig:2}
\begin{minipage}[t]{0.45\linewidth}
\centering
\includegraphics[width=2.5 in]{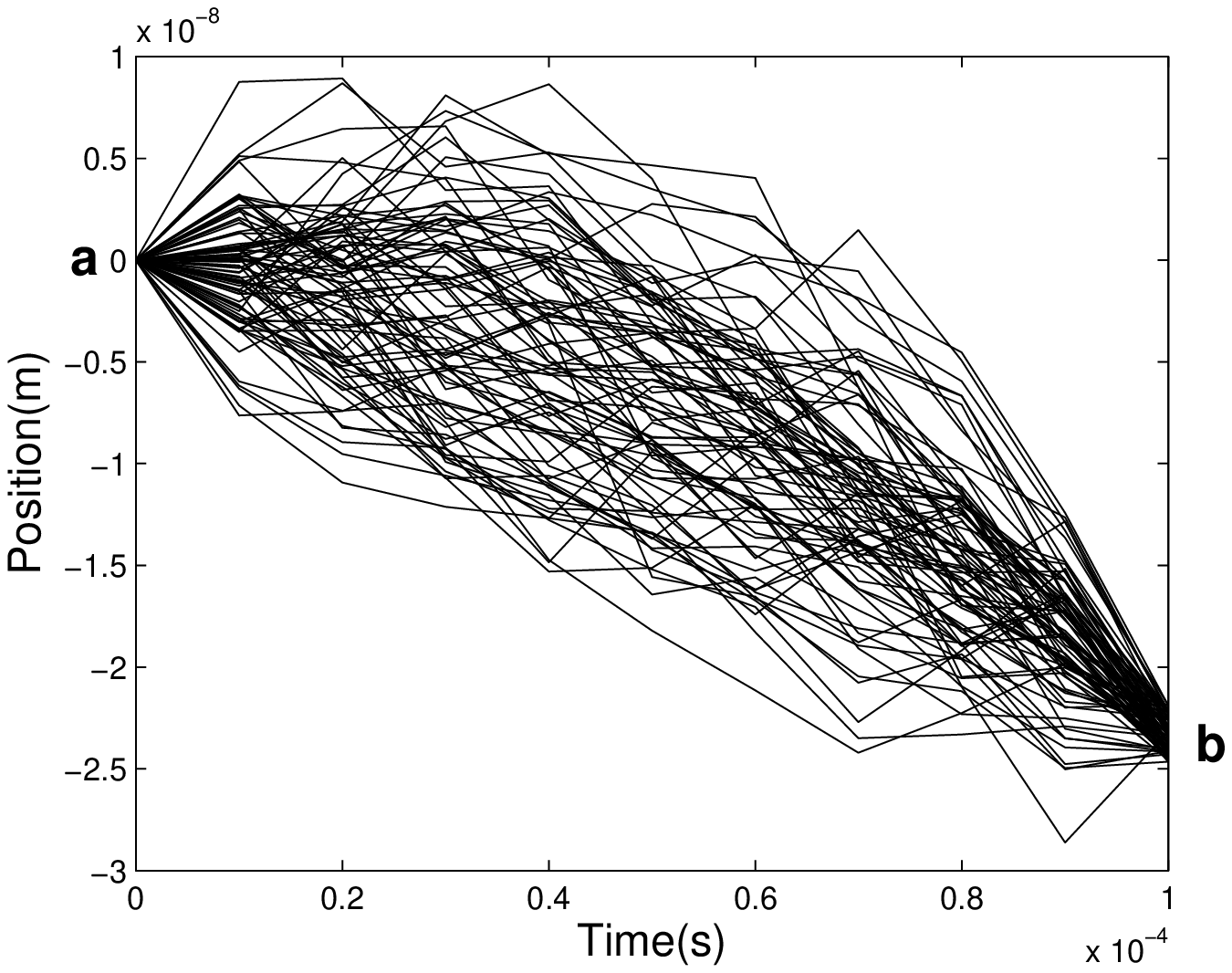}
\end{minipage}%
\begin{minipage}[t]{0.45\linewidth}
\centering
\includegraphics[width=2.5 in]{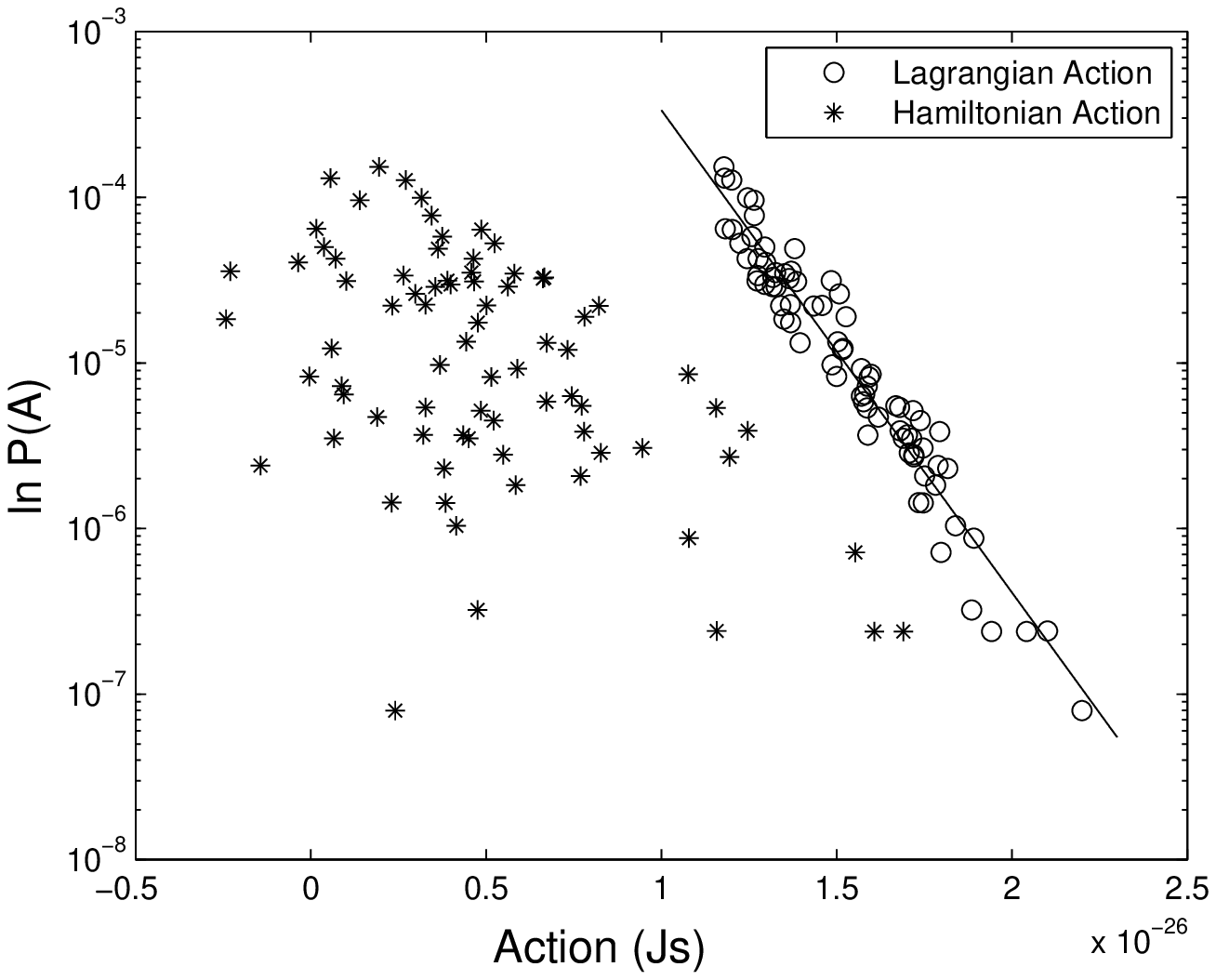}
\end{minipage}
\caption{The result for particles under constant force with potential $V(x)=mgx$. The left panel shows the axial lines of sample paths. In the right panel is depicted the path probability distribution against the Lagrangian (circles) and Hamiltonian (stars) actions. The dependence on the Lagrangian action can be well described by the exponential function of Eq.(\ref{5}) with $\gamma\approx 6.4\times10^{26}$ $J^{-1}s^{-1}$. The straight line is a best fit of the points. There seems no correlation between the path probability and the Hamiltonian action.}
\end{figure}

\subsection{Particles under harmonic force}
The potential of the harmonic force is $V(x)=\frac{1}{2}k x^2$ giving a regular motion $f(t)=A sin(\omega t)$, where $A=1.5\times10^{-8}$ $m$ and $\omega=(k/m)^{1/2}=4.7\times10^4$ $s^{-1}$. The right panel of Figure 3 shows the path probability distribution against actions. As the case of constant force, the path probability distribution decreases exponentially with increasing Lagrangian action with a slope of the straight line $~-7\times10^{26}$ $J^{-1}s^{-1}$. No correlation with the Hamiltonian action can be concluded.

\begin{figure} \label{fig:3}
\begin{minipage}[t]{0.45\linewidth}
\centering
\includegraphics[width=2.5 in]{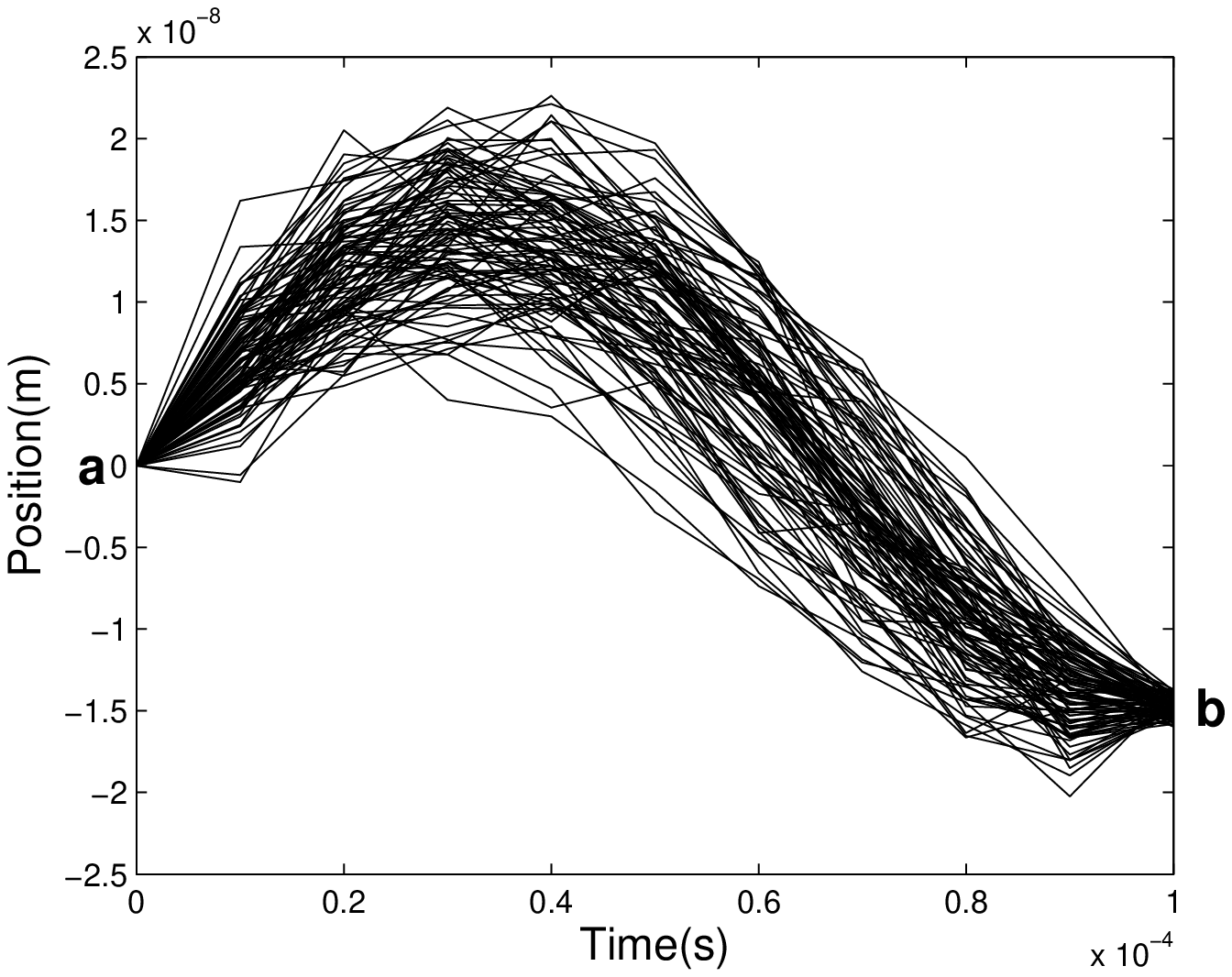}
\end{minipage}%
\begin{minipage}[t]{0.45\linewidth}
\centering
\includegraphics[width=2.5 in]{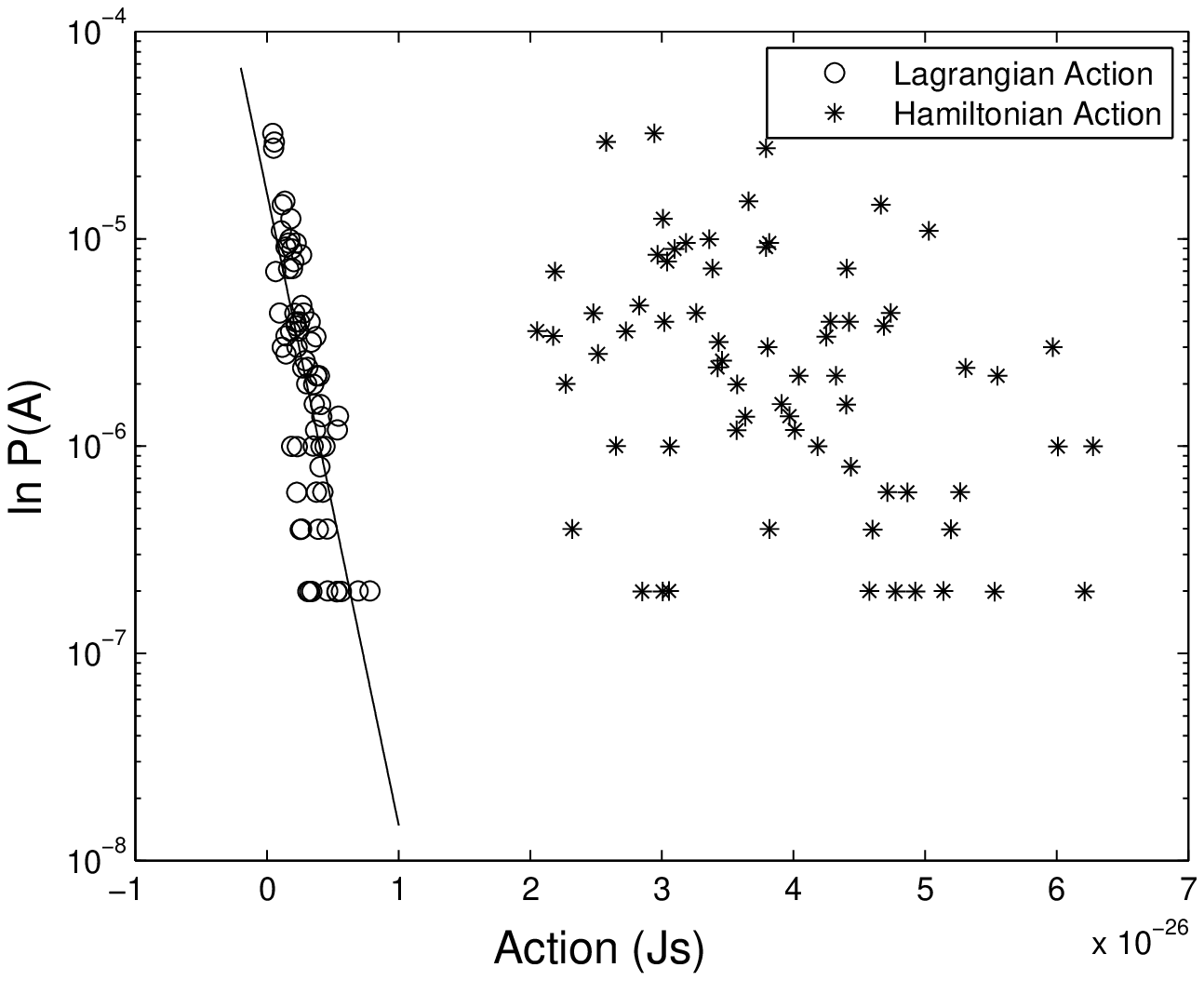}
\end{minipage}
\caption{The result for particles subject to a harmonic force with $V(x)=\frac{1}{2}kx^2$. The left panel shows the axial lines of the sample paths. The right panel shows the path probability distribution against Lagrangian (circles) and Hamiltonian (stars) actions. The dependence on the Lagrangian action can be well described by the exponential function of Eq.(\ref{5}) with $\gamma\approx 7\times10^{26}$ $J^{-1}s^{-1}$. The straight line is a best fit of the points.}
\end{figure}

\subsection{Particles in cubic potential}

To our opinion, the above results with 3 potentials are sufficient to conclude that the path probability can be well described by the exponential function of Lagrangian action. Nevertheless, by curiosity, we also tried two other higher order potentials. The first one is $V(x)=\frac{1}{3}Cx^3$ giving a regular motion $f(t)=-\frac{6m}{C(t_0+t)^2}$ $(t_0=3\times10^{-5}$, $C=200)$. The path probability distributions against the two actions are shown in Figure 4. Eq.(\ref{5}) holds for the Lagrangian action with the coefficient $\gamma\approx 4.5\times10^{26}$ $J^{-1}s^{-1}$.

\begin{figure} \label{fig:4}
\begin{minipage}[t]{0.45\linewidth}
\centering
\includegraphics[width=2.5 in]{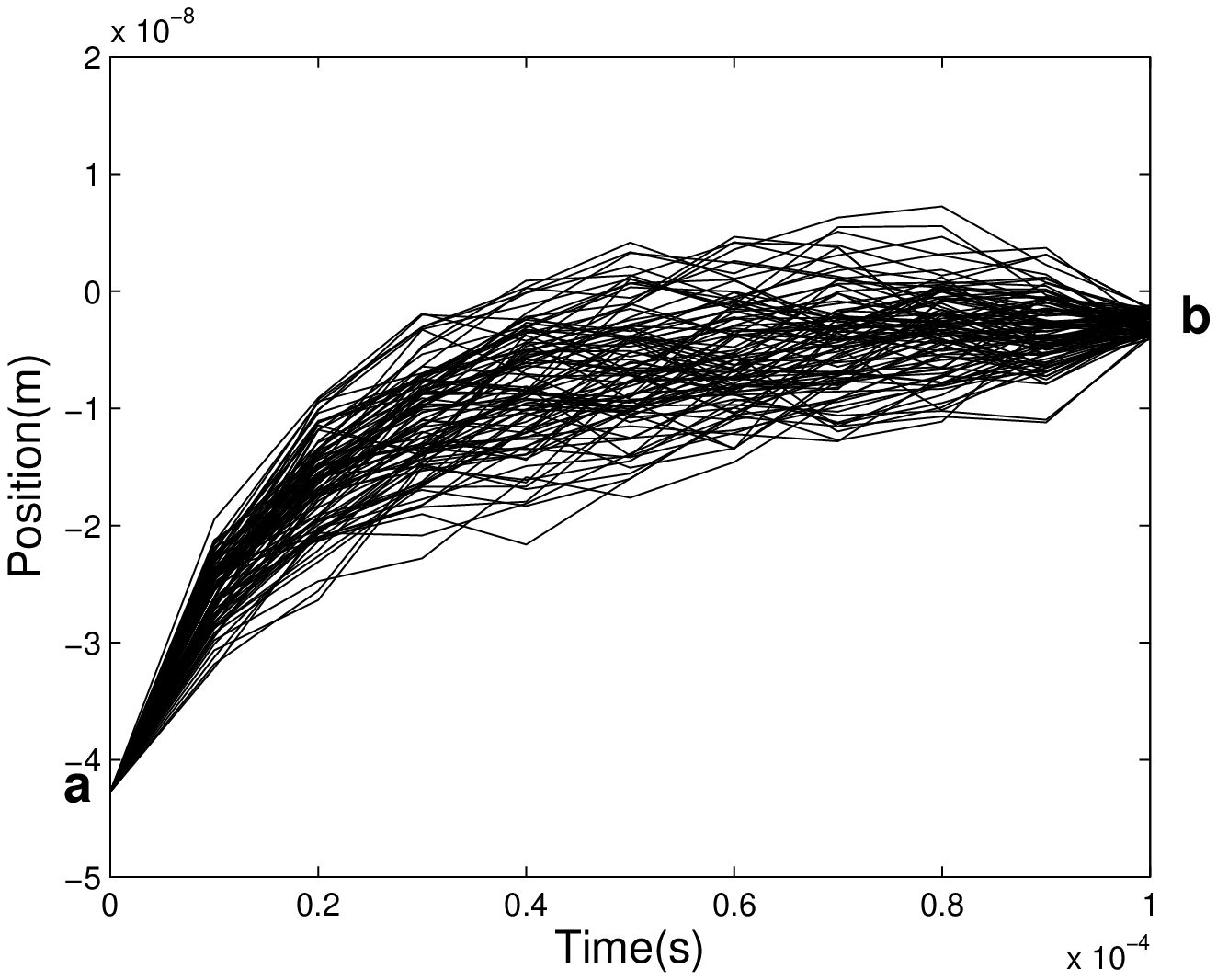}
\end{minipage}%
\begin{minipage}[t]{0.45\linewidth}
\centering
\includegraphics[width=2.5 in]{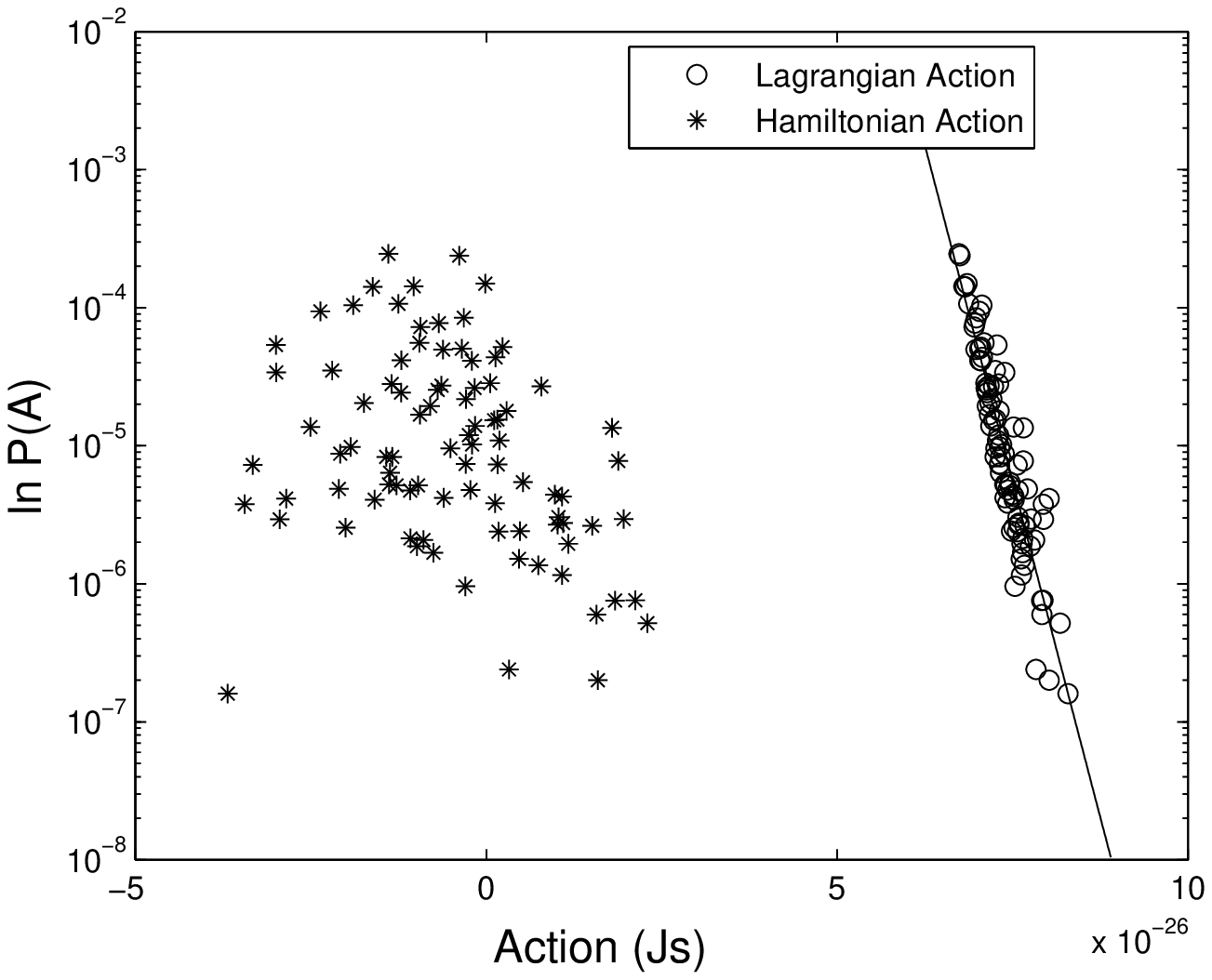}
\end{minipage}
\caption{The result for particles in a cubic potential $V(x)=\frac{1}{3}Cx^3$. The left panel shows that the axial lines of sample paths. The right panel shows the path probability distribution against Lagrangian (circles) and Hamiltonian (stars) actions. The straight line is a best fit of the points with $\gamma\approx 4.5\times10^{26}$ $J^{-1}s^{-1}$ for Eq.(\ref{5}).}
\end{figure}

\subsection{Particles in quartic potential}
For the one-dimensional quartic oscillator, the potential has the form $V(x)=\frac{1}{4}Cx^4$ generating a regular motion which can be approximated by $f(t)\approx A_m sin(\omega t)$ \cite{13,14,15}. Unlike the harmonic potential, the frequency $\omega$ depends on the amplitude $A_m$ in the following way $\omega=\frac{2\pi}{T}\approx(\frac{3C}{4m})^{1/2}A_m=2\times10^{4}$ $s^{-1}$ \cite{13}, where $T$ is the complete cycle period (in the simulation, we have chosen $A_m=1\times10^{-8}$ $m$). The path probability distributions against the two actions are shown in Figure 5. The distribution Eq.(\ref{5}) with the Lagrangian action is confirmed with $\gamma\approx 7.8\times10^{26}$ $J^{-1}s^{-1}$.
\begin{figure} \label{fig:5}
\begin{minipage}[t]{0.45\linewidth}
\centering
\includegraphics[width=2.5 in]{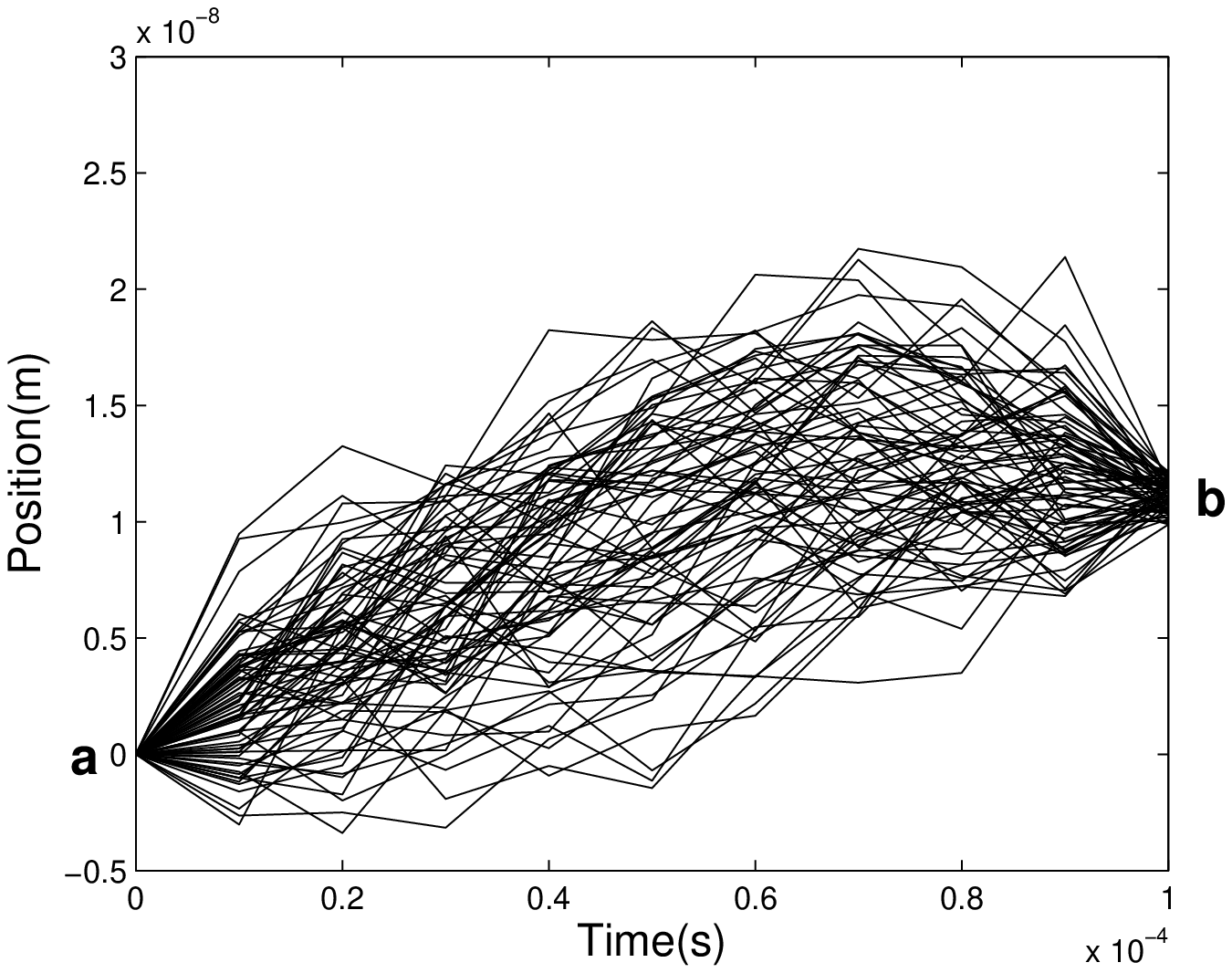}
\end{minipage}%
\begin{minipage}[t]{0.45\linewidth}
\centering
\includegraphics[width=2.5 in]{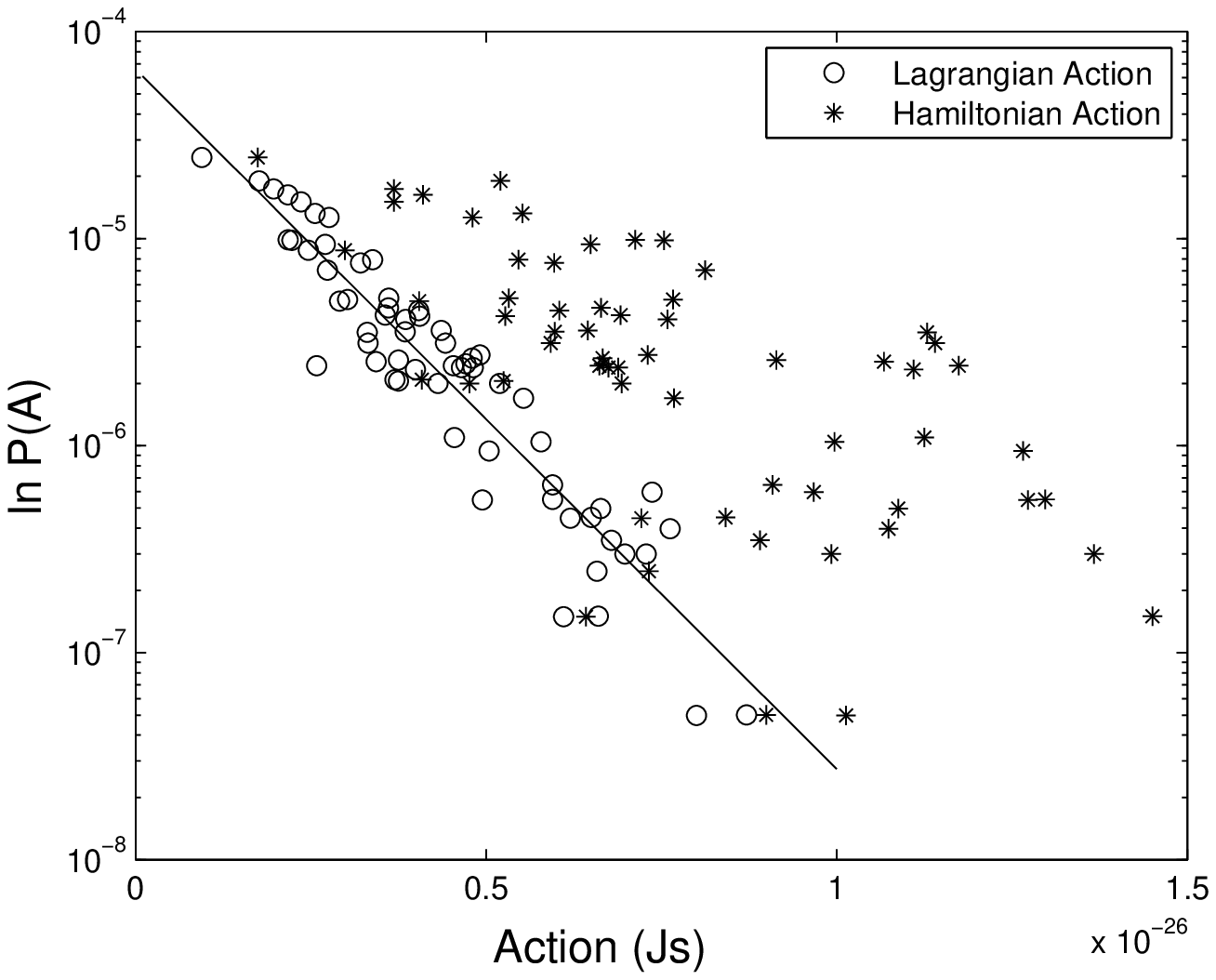}
\end{minipage}
\caption{The result for particles subject to a quartic potential. The left panel shows the axial lines of different sample paths. The right panel shows the path probability distribution against Lagrangian (circles) and Hamiltonian (stars) actions. The straight line is a best fit of the points with $\gamma\approx 7.8\times10^{26}$ $J^{-1}s^{-1}$ for Eq.(\ref{5}).}
\end{figure}

\section{Correlation between path probability and action}
The path probability distributions depicted in Figure 1 to Figure 5 qualitatively confirm an exponential dependence on the Lagrangian action. To our opinion, the reliability of the result are rather remarkable taking into account the mediocre condition of simulation due to the limited computation time which restricts the number of steps of the motion and the minimum thickness of the sample paths. Larger number of steps would make the paths smoother and the calculation of velocity and action more reliable. Smaller thickness of the sample paths would reduce the uncertainty of the probability calculation for given action evaluated along the axial line of a sample path. But larger number of steps and smaller thickness of sample paths will reduce enormously the number of particles arriving at the end point and hence amplifies the uncertainty of the probability calculation. The choice of these two parameters must be optimized according to the computer power. 

The quality of the computation of the probability distribution can be quantitatively estimated by using the correlation function $c(A)$ between $A$ ($A_L$ or $A_H$)$)$ and $-\ln P(A)$. This function is given by
\begin{equation}
c(A)=\frac{\sum_{i=1}^{n}(A_i-<A_i>)[-lnP(A_i)+<lnP(A_i)>]}{\sqrt{[\sum_{i=1}^{n}(A_i-<A_i>)^2][\sum_{i=1}^{n}(-lnP(A_i)+<lnP(A_i)>)^2]}},
\end{equation}
where $<A_i>$ and $<\ln P(A_i)>$ are the means of action $A$ and $-\ln P(A)$ respectively. If $A$ and $-\ln P(A)$ are linearly dependent on each other, $c(A)=1$. The calculated values of $c(A)$ are given in Table 1.

\begin{table}[ht]
\renewcommand{\arraystretch}{1.3}
\caption{Values of the correlation function $c(A)$ between the path probability $-\ln P(A)$ and the Lagrangian action $A_L$ in comparison with the Hamiltonian one $A_H$ for the 5 considered potentials $V(x)$. The values of $c(A_L)$ close to unity confirms a linear correlation between $-\ln P(A)$ and $A_L$. The values of $c(A_H)$ are calculated for comparison. $c(A_H)$ is equal to $c(A_L)$ for free particles due to zero (or constant) potential energy. The fact that the addition of potentials does not significantly change $c(A_L)$ but considerably changes $c(A_H)$ with respect to the free particle values is another element advocating for the universal $A_L$ dependence of the path probability.} % title of Table
\centering
\begin{tabular}{*{4}{p{.2\textwidth}}}
\hline \hline
$V(x)$ & $c(A_L)$ & $c(A_H)$\\
\hline   0 & 0.9865 & 0.9865\\
         $mgx$& 0.9686 & 0.4206\\
         $\frac{1}{2}kx^2$ & 0.9473 & 0.3504\\
         $\frac{1}{3}Cx^3$& 0.9162 & 0.2302\\
         $\frac{1}{4}Cx^4$& 0.9397 & 0.5635\\
\hline\hline
\end{tabular}
\label{table:1} % is used to refer this table in the text
\end{table}

The values of $c(A_L)$ close to unity confirms a linear correlation between $-\ln P(A)$ and $A_L$. It should be noticed that $c(A_H)$ and $c(A_L)$ are equal for free particles due to zero potential energy, and that $c(A_L)$ for different potentials are close to that for free particles while $c(A_H)$ for different potentials are quite different.  This fact that the addition of potentials does not significantly change $c(A_L)$ but does considerably change $c(A_H)$ with respect to the free particle result is another proof of the $A_L$ dependence of the path probability. 

It has been also noticed that $c(A)$ is independent of the time scale $\delta t$ (from $10^{-7}$ to $10^{-3}$ $s$).

\section{Sensitivity of path probability to action}
The decay rate of path probability with increasing action or its sensitivity to action is characterized by the constant $\gamma$. The numerical experiments being performed with different time interval $\delta t$, we noticed that $\gamma$ is $\delta t$ independent. Logically, this sensitivity should be dependent on the randomness of the Gaussian noise. Analysis of the probability distributions reveals that the ratio $\gamma/(1/D)\approx3.2\times10^{14}$ $kg^{-1}$ for free particles, $\gamma/(1/D)\approx3.1\times10^{14}$ $kg^{-1}$ for particles subject to constant force, and $\gamma/(1/D)\approx1.7\times10^{14}$ $kg^{-1}$ with harmonic force. Fig.6. shows the $1/D$ dependence of $\gamma$ for constant force as an example. 

\begin{figure} \label{fig:6}
\centering
\includegraphics[width=3.5 in]{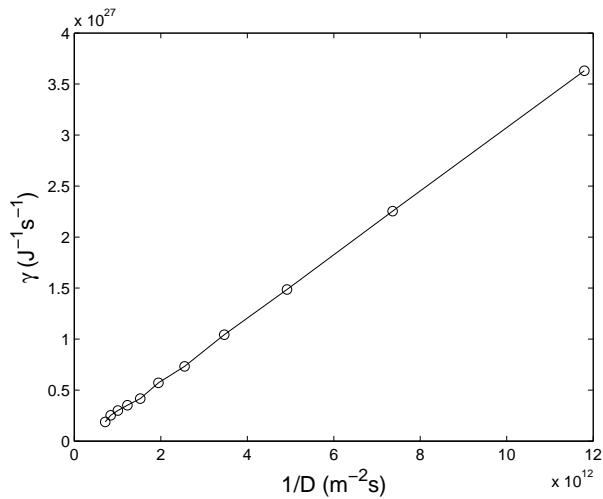}
\caption{$1/D$ dependence of the decay rate $\gamma$ of path probability with action. The increase of $\gamma$ with increasing $1/D$ implies that the stochastic motion is more dispersed around the least action path with more diffusivity. The slope or the ratio $\gamma/(1/D)$ is about $3.1\times10^{14}$ $kg^{-1}$.}
\end{figure}

As expected, $\gamma$ increases with increasing $1/D$, i.e., the stochastic motion is less dispersed around the least action path with decreasing diffusivity $D$. It is worth noticing the linear $1/D$ dependence of $\gamma$ in the range studied here. From theoretical point of view, $\gamma$ should tend to infinity for vanishing $D$. This is the limit case of regular motion of Hamiltonian mechanics. This asymptotic property can also be seen with the uncertainty relation of action given by the standard deviation $\sigma_A\geq\frac{1}{\sqrt{2}\gamma}$\cite{Wang3}. To give an example, when $\gamma=3\times10^{27}$ $(Js)^{-1}$, $\sigma_A\geq 2.4\times10^{-28}$ $Js$. When $\gamma$ is infinity, the least uncertainty of measure of action should be zero, in principle.

\section{Concluding remarks}

To summarize, by numerical simulation of Gaussian stochastic motion of non dissipative systems, we have shown the evidence of the exponential action dependence of path probability. In spite of the uncertainty due to the limited computation time, the computation of the mechanical quantities and the path probability is rigorous and reliable. It is possible to improve the result by more precise computation with longer motion duration, thiner sample paths and more particles. Experimental verification with weakly damped motion can also be expected. 

Apart from the possibility of application to weakly damped motion, the present result seems to provide, by the striking similarity between classical stochastic motion and quantum motion of Hamiltonian systems concerning the choice of paths, an opportunity for a deeper understanding of random motions, and a new angle to see into the random motions and their relationship with non equilibrium statistical mechanics and thermodynamics, in benefiting fully from the technique of path integral already developed in quantum mechanics. An example of this tool borrowing is shown in \cite{Wang4} for the discussion of possible classical uncertainty relations. 

Unlike the Feynman factor $e^{iA/\hbar}$ which is just a mathematical object, $e^{-\gamma A}$ is a real function describing the path probability. This exponential form and the positivity of $\gamma$ imply that the most probable path is just the least action path of classical mechanics, and that when the noise diminishes, more and more paths will shrink into the bundle of least action paths. In the limit case of vanishing noise, all paths collapse on the least action path, the motion recovers the Newtonian dynamics. 

The present result does not mean that the probability for single trajectory necessarily exists. Each path considered here is a tube of thickness $\delta$ and is sufficiently smooth and thin for the instantaneous position and velocity determined along its axial line to be representative for all the trajectories in it. The probability of such a path should tend to zero when $\delta\rightarrow 0$. However, the density of path probability should have a sense and can be defined by $\rho_k=\lim_{\delta\rightarrow 0}\frac{P_k}{\delta^n}$ for any finite $n$, the number of steps of a discrete random process.

Again, we would like to stress that the present results do not apply to the usual Brownian motion or similar overdamped stochastic motions. These motions have been well studied with Langevin, Fokker-Planck and Kolmogorov equations which include friction forces. This work is not at odds with these well established approaches. This is a different angle to address stochastic dynamics. It is our hope that it will be applied to real stochastic dissipative motion. This application needs, first of all, a fundamental extension of the least action principle to dissipative regular motion within classical mechanics. It is unimaginable that the action, being no more a characteristic variable of the paths of regular motion, can come into play when the same motion is perturbed by noise. This extension is another long story, and has been the objective of unremitting efforts of physicists till now\cite{Lanczos,Sieniutycz,Wang1,Wang5}.

\section*{Acknowledgements}
This work was supported by the Region des Pays de la Loire in France under the grant number No. 2007-6088 and No. 2009-09333.

\end{document}